\documentclass[twocolumn,english,aps,prb,superscriptaddress,bibnotes,amsmath,amssymb,floatfix]{revtex4-2}
\usepackage[colorlinks=true,citecolor=blue,linkcolor=magenta]{hyperref}
\usepackage{xcolor}
\usepackage[markup=nocolor, authormarkupposition=left]{changes} 
\usepackage{soul}
\usepackage[utf8]{inputenc}
\usepackage[english]{babel}
\usepackage{amsmath,amsfonts,amssymb}
\usepackage[T1]{fontenc}
\usepackage{url}
\usepackage{epstopdf}
\usepackage{graphicx}
\graphicspath{{./Figures/}}

\begin{document}

% ------------------------------------------ TITLE ---------------------------------
\title{Material-Anisotropy-Driven Topological Optical Lattices on Thin-Film Lithium Niobate}

% ------------------------------------------ AUTHORS ---------------------------------

\author{Siyuan Zhang}
\affiliation{State Key Laboratory of Opto-Electronic Information Acquisition and Protection Technology, School of Physical Sciences, University of Science and Technology of China, Hefei, 230026, China}

\author{Baoqi Shi}
\affiliation{State Key Laboratory of Opto-Electronic Information Acquisition and Protection Technology, School of Physical Sciences, University of Science and Technology of China, Hefei, 230026, China}
%\affiliation{Hefei National Laboratory, University of Science and Technology of China, Hefei, 230088, China}
\affiliation{International Quantum Academy, Shenzhen, 518048, China}

\author{Lei Gui}
\affiliation{State Key Laboratory of Opto-Electronic Information Acquisition and Protection Technology, School of Physical Sciences, University of Science and Technology of China, Hefei, 230026, China}

\author{Xiangle Li}
\affiliation{State Key Laboratory of Opto-Electronic Information Acquisition and Protection Technology, School of Physical Sciences, University of Science and Technology of China, Hefei, 230026, China}

\author{Junna Yao}
\affiliation{State Key Laboratory of Opto-Electronic Information Acquisition and Protection Technology, 
School of Physical Sciences, University of Science and Technology of China, Hefei, 230026, China}

\author{Zhaosheng Chu}
\affiliation{State Key Laboratory of Opto-Electronic Information Acquisition and Protection Technology, School of Physical Sciences, University of Science and Technology of China, Hefei, 230026, China}

\author{Jun Xu}
\affiliation{State Key Laboratory of Opto-Electronic Information Acquisition and Protection Technology, School of Physical Sciences, University of Science and Technology of China, Hefei, 230026, China}

\author{Qiwen Zhan}
\affiliation{Zhejiang Key 
Laboratory of 3D Micro/Nano Fabrication and Characterization, Department of Electronic and Information Engineering, School of Engineering, Westlake University, Hangzhou, 310030, China}

\author{Junqiu Liu}
\affiliation{International Quantum Academy, Shenzhen, 518048, China}
\affiliation{Hefei National Laboratory, University of Science and Technology of China, Hefei, 230088, China}

\author{Anting Wang}
\email[Corresponding author: ]{atwang@ustc.edu.cn}
\affiliation{State Key Laboratory of Opto-Electronic Information Acquisition and Protection Technology, School of Physical Sciences, University of Science and Technology of China, Hefei, 230026, China}

\maketitle

% -------------------------------------------- ABSTRACT ---------------------------
\boldmath
\noindent\textbf{Integrated structured-light sources usually obtain high-dimensional orbital angular momentum (OAM) states by encoding each channel into separate gratings, waveguides or metasurfaces, which ties modal capacity to structural complexity.
Here we show that intrinsic material anisotropy can instead act as a built-in angular-momentum coupler.
In an X-cut thin-film lithium niobate (TFLN) microring vortex emitter, the in-plane optical axis causes a circulating whispering-gallery mode to sample a periodically varying effective index, producing continuous azimuthal phase modulation.
This modulation converts each resonance from a nominal single-charge emitter into a coherent topological sideband lattice with charges $l=l_p+2n$ and Bessel-weighted amplitudes.
Broadband measurements resolve a representative principal-charge series from $l_p=-13$ to $+13$, while additional devices with 100 and 200 GHz free spectral ranges (FSRs) show scalable resonance addressability.
The emitted lattices are reproduced by a forward-calculated Fourier--Bessel model, supported by OAM projection measurements, and exhibit focusing into annular perfect-vortex fields and self-healing after obstruction.
Waveguide-induced circular polarization further adds a vectorial spin--orbit channel.
These results turn TFLN anisotropy from a material constraint into a compact mechanism for resonance-addressed high-dimensional structured-light generation.}
\unboldmath

% ----------------------------------------- MAIN TEXT ----------------------------------
Structured light carrying orbital angular momentum (OAM) has become a central resource for photonics because it provides an unbounded set of orthogonal azimuthal states and enables spatial wavefields that cannot be represented by a single scalar beam mode \cite{Allen1992,Shen2019}.
High-dimensional OAM states are relevant to mode-division multiplexing in optical communications \cite{Willner2015,Wang2022,Ruan2024}, optical computing \cite{Chen2026,Chu2023}, high-dimensional quantum encoding \cite{Erhard2018,Wang2018,Kim2025}, and optical manipulation \cite{Padgett2011,Grier2003,Panda2024}.
For these functions to move from free-space optical tables to deployable photonic systems, the generator must not only emit a chosen OAM order but also address and couple many orders on chip.

Integrated OAM emitters have so far relied mainly on structure-defined angular momentum.
Sidewall-grating microrings can convert a whispering-gallery mode into a vortex beam, but conventional isotropic implementations are naturally organized around a near one-to-one relation between a resonance and a principal topological charge \cite{Cai2012,Zhang2020,Wang2024}.
To access multiple charges or multifunctional vortex states, existing platforms introduce additional structural or operational degrees of freedom, including multiport waveguide coupling and interleaved gratings \cite{Miao2016,Huang2025}, wavelength-dispersive Rowland-circle architectures \cite{Zhang2026}, metasurface-assisted emitters \cite{Gao2025}, and nonlinear frequency-conversion, microcomb or thin-film lithium niobate (TFLN) nonlinear chip--space-interface routes \cite{Chen2024,Liu2024,Li2024,Ahmed2025,Wei2026}.
These advances highlight the importance of coupling spectral, spatial and modal degrees of freedom on chip, but the additional OAM content is usually introduced by engineered scatterers, nonlinear cavity dynamics, external control fields or multiport photonic circuits.
Here the angular-momentum coupling is instead supplied by the material anisotropy itself, so that each individual resonance is converted into a coherent OAM sideband lattice without assigning each topological channel to a separate emitter.

% Figure 1
\begin{figure*}[t!]
\centering
\includegraphics[width=\linewidth]{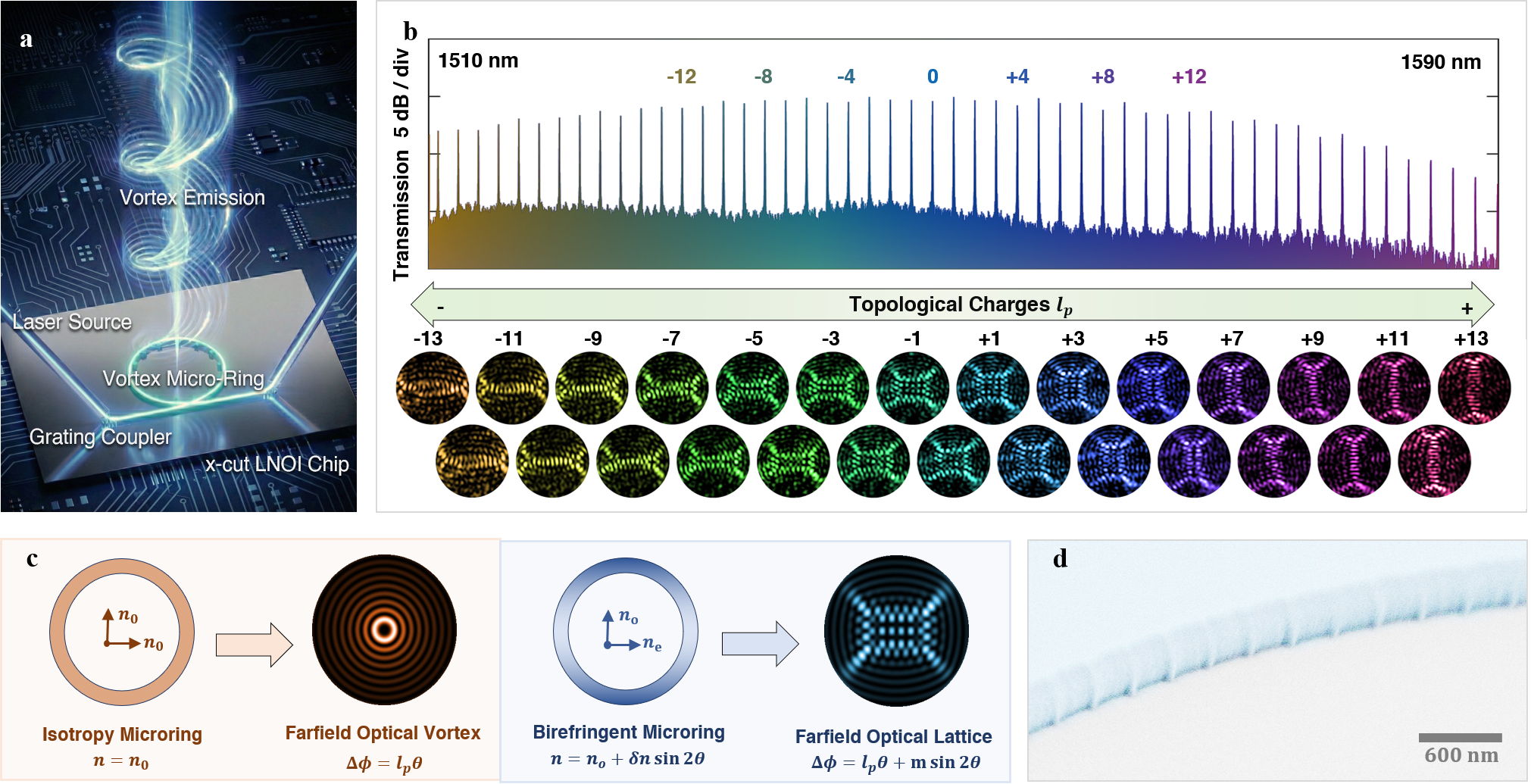}
\caption{
\textbf{Material-anisotropy-driven topological sideband lattice in an X-cut TFLN microring.}
\textbf{a}, Conceptual schematic of the TFLN vortex emitter. A continuous-wave laser is coupled into the bus waveguide through on-chip grating couplers and excites a sidewall-grating microring, which radiates structured light into free space.
\textbf{b}, Measured free-space scattered optical power during a broadband wavelength sweep. The lower image array shows far-field patterns for $l_p=-13$ to $+13$, with labelled odd charges in the upper row and intervening even charges displayed in the lower row. Each resonance emits a lattice-like wavefield rather than a single isolated vortex. Higher-order resonance states and broadband libraries from 100- and 200-GHz free-spectral-range microrings are shown in Supplementary Note 3.
\textbf{c}, Schematic comparison between a conventional isotropic microring vortex emitter, where one resonance predominantly emits one topological charge, and an X-cut TFLN microring, where birefringence-induced azimuthal phase modulation couples the principal charge to evenly spaced OAM sidebands.
\textbf{d}, Scanning electron microscope image of the fabricated device, showing the nanoscale angular gratings etched on the microring sidewall.}
\label{fig:1}
\end{figure*}

X-cut TFLN provides such a route.
Because the crystallographic Z-axis lies in the chip plane, a guided mode circulating around a microring continuously changes its propagation direction relative to the optical axis.
The resulting direction-dependent effective index imposes a periodic phase accumulation along the azimuthal coordinate, so intrinsic birefringence becomes a built-in angular phase modulator.
Although such birefringent phase modulation could in principle be implemented in other anisotropic waveguides, X-cut TFLN is particularly suited to this mechanism: its in-plane crystal axis provides deterministic twofold angular modulation, while the lithium niobate-on-insulator (LNOI) platform supports low-loss nanophotonic confinement and high-Q microring resonances.
The strong Pockels electro-optic response and second-order nonlinearity of lithium niobate also provide routes towards active tuning, high-speed modulation \cite{Boes2023,Zhu2021} and nonlinear wavelength conversion \cite{Wei2026} of the generated sideband lattices.
Thus, LNOI is used here not merely as a birefringent medium, but as an active anisotropic nanophotonic platform.

Here we use this effect to transform a compact microring vortex emitter into a resonance-addressed source of topological OAM sideband lattices.
The key advance is not simply the observation of non-circular far-field patterns, but the conversion of each resonant principal charge into a coherent ladder of coupled charges whose spacing is fixed by the twofold symmetry of the anisotropy, as summarized by the contrast between the conventional isotropic emitter and the X-cut anisotropic emitter in Fig.~\ref{fig:1}c.

% ------------------------------------------ RESULTS ---------------------------------
\section*{Results}
%\par\smallskip\noindent\textbf{Results}

\par\noindent\textbf{Intrinsic anisotropy as an angular phase modulator.}
The device combines a sidewall-grating microring vortex emitter with the in-plane birefringence of X-cut TFLN; the optical layout is shown in Fig.~\ref{fig:1}a, and the fabricated sidewall-grating microring is shown in Fig.~\ref{fig:1}d (see Methods, Sample fabrication and characterization).
The angular grating scatters the circulating whispering-gallery mode into free space, whereas the anisotropic waveguide supplies an additional phase modulation before radiation.
Because the local propagation direction rotates continuously around the ring, the effective index oscillates with twofold angular symmetry.
The accumulated anisotropy-induced phase shift can be written as
\begin{equation}
\Delta\phi(\theta)=m\sin(2\theta),
\label{eq:1}
\end{equation}
where $m$ is the signed modulation amplitude.
With the convention $\delta n_{\text{eff}}=n_{\text{eff},Y}-n_{\text{eff},Z}$, this amplitude is $m=-\pi R\delta n_{\text{eff}}/(2\lambda)$.
Changing the azimuthal-coordinate convention reverses the sign of $m$, which changes relative sideband phases but not the ideal modal power envelope $|J_n(m)|^2$.
A derivation based on the optical indicatrix and guided-mode effective indices is provided in Methods, Theoretical modeling of birefringence-induced phase modulation.

% Figure 2
\begin{figure*}[t!]
\centering
\includegraphics[width=\linewidth]{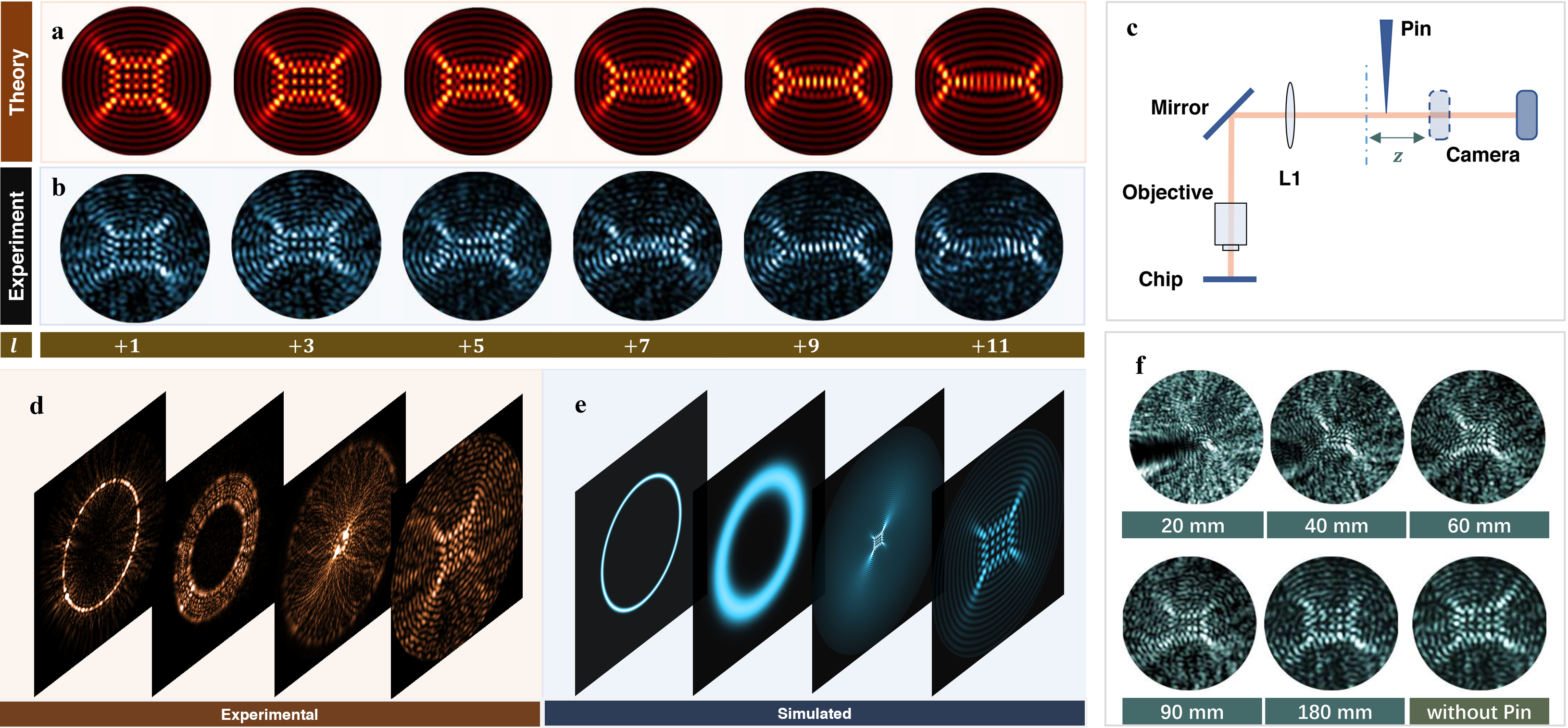}
\caption{
\textbf{Bessel-weighted sideband lattices and propagation dynamics.}
\textbf{a}, Analytical far-field intensity patterns calculated from the Fourier--Bessel model for representative positive principal charges ($l_p=+1,+3,+5,+7,+9,+11$).
\textbf{b}, Corresponding experimentally captured far-field intensity distributions.
\textbf{c}, Optical configuration for the propagation and self-healing experiment. Dashed and solid camera boxes mark axial camera positions; $z$ denotes the distance from the chip image plane (dashed line) formed by the relay lens to the camera, as also used in \textbf{f}. A metallic needle is placed near the image plane as a macroscopic obstacle.
\textbf{d,e}, Longitudinal evolution of the structured wavefield along the propagation axis, comparing experiment (\textbf{d}) and diffraction simulation (\textbf{e}). At the focal plane, the phase-locked sidebands form a narrow annular perfect-vortex field; upon defocusing, the lattice-like morphology reappears.
\textbf{f}, Experimental self-healing sequence showing partial reconstruction of the disrupted lattice-like wavefield after propagation beyond the obstacle.}
\label{fig:2}
\end{figure*}

This angular phase modulation is directly reflected in the broadband response of the fabricated devices.
During the broadband pump-wavelength sweep, the out-of-plane scattered signal shows a sequence of resonantly addressed emission states (Fig.~\ref{fig:1}b).
The main-text series displays principal charges from $l_p=-13$ to $+13$; transmission spectra and $Q$-factor statistics for the measured devices are summarized in Supplementary Note 1, and higher-order states are provided in Supplementary Note 3.
The far-field profiles evolve into lattice-like wavefields with broken continuous rotational symmetry, indicating that each resonance radiates a coherent superposition of OAM components rather than a single isolated charge.
The same measurement protocol applied to microrings with 100 GHz and 200 GHz free spectral ranges (FSRs) shows that reducing the FSR increases the number of addressable resonant states within a fixed wavelength window, providing a device-level handle for scaling the principal-charge library.
This comparison also shows that resonator design controls more than the resonance density.
Because the modulation depth satisfies $m=-\pi R\delta n_{\mathrm{eff}}/(2\lambda)$, changing the microring radius modifies both the spectral density of addressable principal charges and the Bessel-weighted bandwidth of the anisotropy-induced sidebands.
The 100-GHz device therefore extends the wavelength-addressed charge library, whereas the 200-GHz device provides a more sparsely sampled but otherwise equivalent manifestation of the same sideband-generation mechanism (Supplementary Note 3).
This radius/FSR dependence is important for scaling the platform because it separates two linked design knobs: the number of resonances available in a target wavelength band and the modal width of the sideband lattice emitted from each resonance.

\begin{figure*}[t!]
\centering
\includegraphics[width=\linewidth]{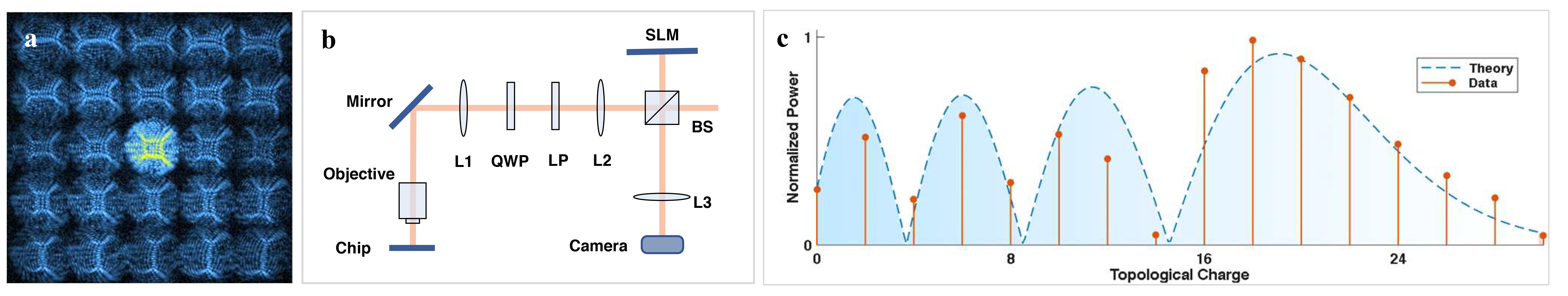}
\caption{
\textbf{Projection-based modal check of the anisotropy-induced OAM sidebands.}
\textbf{a}, Single-frame visualization of selected projection channels generated with a $5\times5$ Dammann vortex grating over the charge range from $l=-12$ to $l=+12$.
\textbf{b}, Schematic of the spatial-light-modulator (SLM)-based projection setup.
\textbf{c}, Projection-derived modal response obtained by scanning a single helical phase mask $e^{i l_{\mathrm{SLM}}\theta}$ and recording the on-axis Fourier-plane intensity. With the sign convention used here, the plotted physical charge is $l=-l_{\mathrm{SLM}}$. The response peaks at the expected even-order sidebands and provides an auxiliary check of the selection rule $l=l_p+2n$.}
\label{fig:3}
\end{figure*}

\par\smallskip\noindent\textbf{Bessel-weighted topological sideband lattices.}
The lattice-like morphology follows from the angular spectrum imposed by Eq.~\ref{eq:1}.
The out-coupled annular source field contains the principal azimuthal phase $e^{i l_p\theta}$ and the birefringence-induced modulation $e^{i m\sin(2\theta)}$.
Using the Jacobi--Anger expansion, the periodic modulation decomposes into even-order angular harmonics, so the principal charge is coupled to sidebands $l=l_p+2n$.
After the Fourier transform of the annular source field, the emitted scalar field can be written as the Fourier--Bessel form derived in Methods, Far-field wavefield derivation:
\begin{equation}
U_{m,l_p}(r,\theta)=e^{i l_p\theta}\sum_{n=-\infty}^{+\infty}(-1)^nJ_n(m)J_{l_p+2n}(k_r r)e^{i2n\theta},
\label{eq:2}
\end{equation}
where $k_r$ is the transverse spatial-frequency scale set by the Fourier mapping of the microring radius.
The coefficient $J_n(m)$ gives the complex amplitude of the $n$-th anisotropy-induced sideband, $J_{l_p+2n}(k_r r)$ describes its radial profile, and the factor $(-1)^n$ contributes only to relative phase.
Thus the ideal modal power envelope is governed by $|J_n(m)|^2$, whereas the measured intensity pattern arises from coherent spatial interference among the coupled charges.

We compared this forward model with the measured far-field libraries over a representative range of principal charges, with the calculated patterns shown in Fig.~\ref{fig:2}a and the corresponding experimental patterns shown in Fig.~\ref{fig:2}b.
The model reproduces the main lattice-like morphologies without using the images to fit the modulation depth.
Instead, finite-element eigenmode simulations of the fabricated waveguide cross-section, using established lithium-niobate Sellmeier data \cite{Zelmon1997}, give $n_{\text{eff},Y}$ and $n_{\text{eff},Z}$, from which we obtain $|m|=5.80$ at $\lambda=1550$ nm (Supplementary Note 2).
The wavelength dependence of $m$ across 1510--1590 nm is moderate, producing smooth changes in the Bessel envelope rather than a qualitative change in the lattice formation.
This agreement supports the interpretation that the observed patterns originate from material-anisotropy-induced angular momentum coupling.
This distinction is important because the sidewall grating is still the element that extracts the circulating mode into free space, whereas the anisotropic waveguide determines the additional angular spectrum carried by that extracted field.
In an isotropic microring with the same grating order, the dominant selection rule would assign one resonance to one principal charge; here the same type of out-coupler samples a source field that has already acquired the birefringent phase $e^{im\sin(2\theta)}$.
The model therefore separates the role of the grating as an out-coupler from the role of X-cut TFLN as the angular-momentum-coupling medium.

\par\smallskip\noindent\textbf{Propagation dynamics of phase-locked sidebands.}
Because all coupled sidebands are emitted from the same annular microring aperture, the relay microscope images this common source to a fixed ring radius set by the microring geometry and optical magnification rather than by the topological charge.
The focused field therefore forms an annular perfect-vortex field, or equivalently a phase-locked superposition of perfect-vortex modes sharing the same radius.
Experimentally, the lattice-like wavefield collapses near focus into a narrow annular intensity distribution (Fig.~\ref{fig:2}d), and the corresponding diffraction simulation reproduces the same focal transformation and defocused recovery (Fig.~\ref{fig:2}e).
This reversible transformation advances the result from a modal OAM spectrum to a propagating structured wavefield whose spatial form can be optically converted.
Away from the focal plane, the same sidebands interfere again and reconstruct the characteristic lattice-like multi-lobed morphology.

We further tested whether this structured wavefield retains the self-healing behavior associated with Bessel-like angular spectra.
A metallic needle with a maximum diameter of 0.8 mm was placed near the relayed focal plane to perturb the field (Fig.~\ref{fig:2}c).
Immediately after obstruction, the beam profile was strongly ruptured; after propagation, the disrupted lattice-like morphology partially reconstructed (Fig.~\ref{fig:2}f).
Here self-healing refers to recovery of the characteristic lattice geometry associated with the phase-locked OAM sideband structure, not to complete restoration of the original optical field.
The recovery is consistent with inward-propagating conical wave-vector components refilling the shadowed region from unobstructed portions of the beam.

\par\smallskip\noindent\textbf{Modal and vectorial checks.}
We used spatial-light-modulator (SLM)-based projection measurements to check the predicted modal structure of selected emitted states (Fig.~\ref{fig:3}b).
A programmable $5\times5$ Dammann vortex grating \cite{Fu2016} provides a compact single-frame visualization over a central charge range from $l=-12$ to $+12$ (Fig.~\ref{fig:3}a; Supplementary Note 6), but this multiplexed image is not used as a complete quantitative OAM decomposition.
For the projection-derived response in Fig.~\ref{fig:3}c, a single helical phase mask $e^{i l_{\mathrm{SLM}}\theta}$ was scanned and the on-axis Fourier-plane intensity was recorded after background subtraction.
An incident component $e^{i l\theta}$ is converted to $e^{i(l+l_{\mathrm{SLM}})\theta}$, so the on-axis signal is maximized when $l=-l_{\mathrm{SLM}}$.
The measured response shows peaks at the expected sideband orders, supporting the even-order coupling rule.
Because the readout is a finite-aperture scalar projection affected by SLM pixelation, beam centring, background subtraction and radial-mode mismatch, it should be regarded as a modal consistency check rather than a full phase-resolved decomposition.

The same TFLN waveguide also introduces a vectorial emission channel.
For the fundamental quasi-transverse-electric (quasi-TE) whispering-gallery mode, the dominant in-plane field component is accompanied by weaker longitudinal and out-of-plane components whose relative phase can approach quadrature under strong confinement.
When this hybrid guided mode is sampled by the sidewall grating, the radiated field contains measurable circular-polarization components.
Spatially resolved Stokes polarimetry, cross-sectional mode simulations and directional-pumping experiments are presented in Supplementary Notes 4 and 5.
Upon reversing the input port, the intracavity propagation direction is reversed, which flips both the grating-defined phase gradient and the handedness of the circular-polarization component; the resulting far-field pattern is consistent with phase conjugation of the emitted wavefield.
This vectorial response provides an additional spin--orbit degree of freedom \cite{Zhu2014}, but the primary high-dimensional OAM generation mechanism remains the anisotropy-induced scalar phase modulation described above.

% ------------------------------------------ DISCUSSION ---------------------------------
\section*{Discussion}
%\noindent\textbf{Discussion and Conclusion.}
We have demonstrated a material-anisotropy-driven mechanism for high-dimensional structured-light generation in an integrated X-cut TFLN microring.
The central result is that intrinsic birefringence, usually treated as a platform parameter that must be managed, can be used as a continuous angular phase modulator that couples a resonant principal charge to a Bessel-weighted lattice of OAM sidebands.
This changes the design logic of chip-scale OAM generation: rather than assigning each topological channel to an independently engineered structure, a single compact resonator uses its material anisotropy to generate a phase-locked ladder of coupled charges at each resonance.

The experiments connect this mechanism across spectral, spatial, modal and propagation-domain evidence.
Broadband resonance measurements show addressable principal-charge series in the telecom band, and the 100- and 200-GHz-FSR devices illustrate how the number of accessible states can be scaled through the resonator design.
The measured lattice-like far fields are captured by a forward-calculated Fourier--Bessel model, while projection measurements check the predicted even-order sideband rule.
Propagation and obstruction experiments further show that the generated sideband lattices can transform into narrow annular perfect-vortex fields near focus and partially self-heal after macroscopic perturbation.
Together, these observations show that the anisotropy-induced sidebands constitute a propagating and optically transformable structured wavefield, rather than only a static OAM spectrum.
The observed circular-polarization component adds a vectorial spin--orbit channel, although it is not required to explain the scalar sideband lattice.

These observations place the device between two active directions in integrated vortex photonics. It retains the compact resonance-addressed form of microring vortex emitters, but replaces the usual structure-defined one-resonance--one-charge relation with a material-symmetry-defined sideband ladder. It also offers a passive counterpart to vortex microcomb and reconfigurable OAM platforms, because the sideband set is generated within each resonance by the intrinsic anisotropic phase of the waveguide rather than by nonlinear comb dynamics or externally programmed circuits.

Several limitations and design opportunities define the next stage of the platform.
In the present devices, the sideband envelope is fixed mainly by the birefringence-induced modulation depth $m$ and therefore follows a Bessel-weighted distribution.
Future devices could combine the intrinsic phase modulation with engineered grating amplitudes, aperiodic scatterer distributions or active electro-optic tuning to tailor the sideband weights and dynamically switch the emitted wavefields \cite{Boes2023,Zhu2021}.
The free-space extraction efficiency is governed by the sidewall-grating strength, bus--ring coupling condition and vertical radiation asymmetry, and can be further improved by optimizing these structural parameters without changing the material-anisotropy-driven sideband-generation mechanism.
The concept should also be transferable to other anisotropic photonic platforms, but the resulting sideband spacing, modulation depth and accessible modal bandwidth will depend on crystal symmetry, optical-axis orientation, waveguide cross-section and resonator radius.
In X-cut TFLN, the in-plane twofold birefringence naturally selects the even-order ladder $l=l_p+2n$; other crystal cuts or anisotropic materials could, in principle, generate different angular selection rules if their effective-index modulation contains different Fourier harmonics.
This provides a general design principle for using material symmetry as a modal-coupling resource, rather than treating anisotropy only as a perturbation to be compensated.
Together, these results establish X-cut TFLN as a compact anisotropic platform for resonance-addressed high-dimensional structured light, with potential relevance to mode-division multiplexing, optical manipulation, structured illumination and high-dimensional photonic interfaces.

%----------------------------------METHODS----------------------------------
\section*{Methods}
%\par\smallskip\noindent\textbf{Methods}

\noindent\textbf{Theoretical modeling of birefringence-induced phase modulation.}
As schematically indicated in Fig.~\ref{fig:1}c, the intrinsic birefringence of X-cut TFLN introduces an azimuthally varying effective index in the microring resonator.
Because the crystallographic Z-axis of an X-cut lithium niobate film lies in the device plane, the local propagation direction of the guided mode continuously rotates with respect to this optical axis as light circulates around the ring.
This directional dependence leads to a periodic phase accumulation along the azimuthal coordinate and therefore acts as a built-in phase modulation mechanism.

The physical origin of this angular index variation can be described using the bulk index ellipsoid, or optical indicatrix, which captures the direction-dependent refractive index of an anisotropic crystal.
For an idealized bulk X-cut lithium-niobate medium, assuming that the in-plane propagation direction forms an angle $\theta$ with respect to the crystallographic Z-axis, the refractive index $n_\theta$ satisfies the ellipsoidal relation:
\begin{equation}
\frac{n_\theta^2 \sin^2\theta}{n_o^2} + \frac{n_\theta^2 \cos^2\theta}{n_e^2} = 1
\label{eq:4}
\end{equation}
where $n_o$ and $n_e$ are the ordinary and extraordinary refractive indices, respectively.
For weak-to-moderate birefringence, this angular dependence can be written to leading order as a second-harmonic modulation \cite{Herrera2021}:
\begin{equation}
n_\theta = \frac{n_o + n_e}{2} - \frac{n_o - n_e}{2} \cos(2\theta)
\label{eq:5}
\end{equation}

In the actual microring waveguide, the guided-mode effective index is additionally affected by waveguide geometry and modal confinement.
Nevertheless, the same twofold rotational symmetry leads to an approximately sinusoidal variation of the effective index with the azimuthal angle, which we parameterize as:
\begin{equation}
n_{\text{eff}}(\theta) = \frac{n_{\text{eff},Y} + n_{\text{eff},Z}}{2} - \frac{n_{\text{eff},Y} - n_{\text{eff},Z}}{2} \cos(2\theta)
\label{eq:6}
\end{equation}
Here, $n_{\text{eff},Y}$ and $n_{\text{eff},Z}$ denote the simulated effective indices of the guided mode when the local propagation direction is aligned with the crystallographic Y and Z axes, respectively.
This formulation separates the material origin of the anisotropy from the waveguide-specific effective-index response.

The total phase accumulation $\phi$ over a propagation path in the microring is the integral of the local wave vector along the azimuthal direction:
\begin{equation}
\phi(\theta) = \int_0^\theta k_0 n_{\text{eff}}(\theta') R d\theta',
\label{eq:7}
\end{equation}
where $k_0=2\pi/\lambda$ is the free-space wavenumber and $R$ is the microring radius.
By substituting Eq.~\ref{eq:6} into Eq.~\ref{eq:7}, the total phase can be separated into a linear propagation term and an oscillatory term induced by the birefringence.
The anisotropy-induced additional phase shift $\Delta\phi(\theta)$ is then obtained as:
\begin{equation}
\Delta\phi(\theta) = -\frac{\pi R \delta n_{\text{eff}}}{2\lambda} \sin(2\theta),
\label{eq:8}
\end{equation}
where $\delta n_{\text{eff}} = n_{\text{eff},Y} - n_{\text{eff},Z}$, $R$ is the microring radius, and $\lambda$ is the operating wavelength.
The minus sign follows from the above definition of $\delta n_{\text{eff}}$ and the chosen azimuthal coordinate.
Reversing the coordinate direction or the crystal-axis convention changes this sign, which modifies the relative sideband phases but does not affect the ideal modal power envelope.

We define the signed birefringence-induced phase modulation depth $m$ as:
\begin{equation}
m = -\frac{\pi R \delta n_{\text{eff}}}{2\lambda}.
\label{eq:9}
\end{equation}
Consequently, the azimuthal phase modulation imparted onto the emitted wavefield is represented by the term $e^{i m\sin(2\theta)}$.
Through the Jacobi--Anger expansion, this periodic phase modulation couples the principal OAM mode to even-order topological sidebands, with an ideal sideband power envelope described by the Bessel weights $|J_n(m)|^2$.

\par\smallskip\noindent\textbf{Far-field wavefield derivation.}
To elucidate the origin of the far-field expression presented in the main text, we derive an analytical scalar Fourier--Bessel representation for the radiated field.
The derivation treats the emitted field as a monochromatic scalar field and assumes a narrow annular source with an approximately uniform azimuthal scattering amplitude.
Finite-aperture effects, vectorial corrections, and possible azimuthal nonuniformity of the grating scattering strength are neglected.
Within these assumptions, the model captures the topological OAM-sideband lattice and the resulting spatial interference responsible for the lattice-like far-field morphology.

In the transverse plane, each angular component of the monochromatic scalar field satisfies the Helmholtz equation:
\begin{equation}
(\nabla_\perp^2 + k_r^2) U(r,\theta) = 0,
\label{eq:10}
\end{equation}
where \(r\) and \(\theta\) denote the radial and azimuthal coordinates in the Fourier plane.
For the Fourier-plane description used to model the lattice-like far-field patterns, the radial spatial-frequency scale is set by the Fourier mapping of the microring radius:
\begin{equation}
k_r = k_0 \frac{R}{f} = \frac{2\pi}{\lambda} \frac{R}{f},
\label{eq:11}
\end{equation}
where $R$ is the microring radius, $f$ is the objective focal length, and $k_0$ is the free-space wave vector.
The annular source field immediately after out-coupling can be idealized as:
\begin{equation}
U_s(\rho,\theta') \propto \delta(\rho-R)e^{i \left(l_p\theta'+m\sin(2\theta')\right)},
\label{eq:12}
\end{equation}
where \(\rho\) and \(\theta'\) are the radial and azimuthal coordinates in the source plane, $l_p$ is the principal topological charge, and $m$ is the signed birefringence-induced phase modulation depth.
A regular Fourier--Bessel basis for the transverse field can then be written as:
\begin{equation}
U(r,\theta) = \sum_{l=-\infty}^{\infty} A_l J_l(k_r r) e^{i l\theta},
\label{eq:13}
\end{equation}
Here, $l$ is an integer representing the azimuthal quantum number, $J_l(k_r r)$ is the Bessel function of the first kind of order $l$, and $A_l$ is the complex amplitude for each azimuthal component.
The Bessel-function radial profile captures the annular spatial structure associated with the Fourier transform of a ring-like source and provides a scalar description of the observed topological self-healing behavior after obstruction.

The angular phase of the modulated source field is:
\begin{equation}
e^{i \left(l_p\theta'+m\sin(2\theta')\right)}
=
e^{i l_p\theta'}e^{i m\sin(2\theta')}.
\label{eq:14}
\end{equation}
To expand the periodic modulation term, we employ the Jacobi--Anger expansion:
\begin{equation}
e^{i m\sin(2\theta')} = \sum_{n=-\infty}^{\infty} J_n(m) e^{i 2n\theta'}.
\label{eq:15}
\end{equation}
Thus, the angular part of the source field becomes:
\begin{equation}
e^{i \left(l_p\theta'+m\sin(2\theta')\right)}
=
\sum_{n=-\infty}^{\infty} J_n(m)e^{i (l_p+2n)\theta'}.
\label{eq:16}
\end{equation}
This expression directly gives the ideal scalar selection rule: the principal topological charge \(l_p\) is coupled to sidebands with azimuthal indices \(l=l_p+2n\).
The even-order spacing originates from the twofold angular symmetry of the birefringence-induced phase modulation.

The Fourier-plane amplitude is proportional to the two-dimensional Fourier transform of the annular source field.
Using the plane-wave expansion:
\begin{equation}
e^{-i k_r r\cos(\theta'-\theta)}
=
\sum_{s=-\infty}^{\infty}
(-i)^s J_s(k_r r)e^{i s(\theta'-\theta)},
\label{eq:17}
\end{equation}
the scalar field in the Fourier plane can be written as:
\begin{equation}
U(r,\theta)
\propto
\int_0^{2\pi}
e^{i \left(l_p\theta'+m\sin(2\theta')\right)}
e^{-i k_r r\cos(\theta'-\theta)}
d\theta' .
\label{eq:18}
\end{equation}
For each angular component with \(L=l_p+2n\), the angular integral satisfies:
\begin{equation}
\int_0^{2\pi} e^{i L\theta'}e^{-i k_r r\cos(\theta'-\theta)}d\theta'
=
2\pi(-i)^L J_L(k_r r)e^{i L\theta}.
\label{eq:19}
\end{equation}
Substituting Eq.~(\ref{eq:16}) into Eq.~(\ref{eq:18}) and using Eq.~(\ref{eq:19}) gives:
\begin{equation}
U_{m,l_p}(r,\theta)
\propto
e^{i l_p\theta}
\sum_{n=-\infty}^{+\infty}
(-1)^n J_n(m) J_{l_p+2n}(k_r r)e^{i 2n\theta}.
\label{eq:20}
\end{equation}
In obtaining Eq.~(\ref{eq:20}), the global factor \(2\pi(-i)^{l_p}\) has been omitted because it does not affect the intensity distribution or modal power spectrum.
This expression is equivalent to Eq.~(\ref{eq:2}) in the main text up to an irrelevant global phase and normalization factor.
It shows that the emitted wavefield is a coherent superposition of the principal mode \(l_p\) and its coupled sidebands \(l_p\pm2, l_p\pm4,\dots\).
Within the ideal scalar model, the complex amplitude of each sideband is weighted by \(J_n(m)\), and the corresponding modal power envelope scales as \(|J_n(m)|^2\).
The spatial interference among these Bessel-weighted sidebands explains the formation of the lattice-like far-field wavefields observed experimentally.
The perfect-vortex transformation discussed in Fig.~\ref{fig:2}d,e is associated with relay imaging and scalar propagation of the same annular source; at the image plane, the ring radius is fixed by the microring aperture and optical magnification rather than by the OAM charge.

\par\smallskip\noindent\textbf{Sample fabrication and characterization.}
The fabrication of the TFLN chip commenced with a surface cleaning protocol to remove particulate and organic contaminants.
The sample was sequentially subjected to acoustic cleaning, piranha solution treatment, and O$_2$ plasma exposure.
The substrate was then primed with Surpass 3000 adhesion promoter to improve resist adhesion.
A layer of hydrogen silsesquioxane (HSQ, FOx-16) electron-beam resist was spin-coated at 4000 RPM.
Device patterning was performed using a high-resolution electron-beam lithography (EBL) system (JEOL JBX-6300FS).
The defined patterns were subsequently transferred into the lithium niobate layer by inductively coupled plasma (ICP) etching (Sentech SI500) with an Ar/Cl$_2$ chemistry.
After etching, the chip was treated with an SC-1 solution to remove redeposited lithium-niobate residues, followed by a hydrofluoric acid (HF) dip to strip the remaining resist.
Finally, a 30-nm SiO$_2$ cladding layer was deposited by ICP-powered plasma-enhanced chemical vapor deposition (ICP-PECVD) using another Sentech SI500 system.

A key fabrication consideration is the dimensional control of the nanoscale angular gratings on the microring sidewall.
In the EBL mask design, the grating perturbations were defined with lateral dimensions of 30$\times$30 nm$^2$ in the radial and azimuthal directions, respectively.
It should be noted that these values refer to the mask-defined perturbation dimensions rather than the final three-dimensional grating profile after etching.
Corrugated sidewall gratings have recently been implemented in TFLN straight waveguides for Bragg filtering \cite{Baghban2017} or slow-light generation \cite{Li2025,Prencipe2021}.
In those devices, the spatial modulation amplitudes are typically much larger, ranging from approximately 120 nm to over 600 nm, to ensure strong intra-waveguide mode coupling.
By contrast, the present free-space vortex emitter requires much weaker azimuthal perturbations to balance out-of-plane radiation with the preservation of the cavity resonance.
At this dimensional scale, the final grating morphology is sensitive to the ion-milling process.
As observed in the scanning electron microscope (SEM) image in Fig.~\ref{fig:1}d, the fabricated sidewall perturbations tend to exhibit a three-dimensional tapered profile, with the grating features becoming smoother toward the upper part of the waveguide.
Through optical parameter sweeping, we found that larger mask-defined perturbations, such as 50$\times$50 nm$^2$, introduce excessive radiative loading, leading to a pronounced reduction of the cavity $Q$ factor and degradation of the emitted-beam morphology.
Conversely, smaller perturbations reduce the extracted free-space power.
The 30$\times$30 nm$^2$ mask-defined perturbation was therefore selected as an empirical compromise between sufficient free-space out-coupling and acceptable resonator quality for high-dimensional wavefield generation.

For optical characterization, the devices were pumped using a tunable continuous-wave laser source (ANDO AQ4321D and Agilent 81642B) with an incident power of 1 mW.
The emission wavelength was monitored using an optical wavelength meter (Yokogawa AQ6151B), and the input pump power was calibrated using a power meter (Thorlabs S145C) in conjunction with a 99:1 fiber coupler.
Light was coupled into the on-chip waveguides via the fabricated grating couplers, with a measured coupling efficiency of approximately 6 dB per facet.
Broadband spectral and far-field measurements were performed by sweeping the tunable laser from 1510 to 1590 nm.
Both the 200 GHz and 100 GHz FSR microring devices were characterized using the same measurement configuration, allowing their resonance-addressed far-field emission libraries to be compared over the same wavelength window.

The out-of-plane emission from the TFLN vortex microring was collected using a 20$\times$ long-working-distance objective lens with numerical aperture (NA) = 0.2.
The collected beam was relayed either to an InGaAs CMOS near-infrared camera (Allied Vision Alvium 1800 U-130 VSWIR, equipped with a Sony IMX990 sensor) for far-field intensity imaging or to an amplified InGaAs photodetector (Thorlabs PDA10CS2) for wavelength-swept intensity characterization.
The photodetector output was recorded and analyzed using an oscilloscope (Keysight MXR204A, up to 2 GHz).
For projection-based modal analysis, the collection path incorporated a phase-only spatial light modulator (SLM, Thorlabs Exulus-HD4).
An established Dammann vortex grating was used only for single-frame visualization of selected OAM projection channels, while quantitative projection responses were obtained by scanning single helical phase masks $e^{i l_{\mathrm{SLM}}\theta}$ and recording the on-axis Fourier-plane intensity.
Under this sign convention, the measured spectral axis corresponds to the physical topological charge $l=-l_{\mathrm{SLM}}$.

%------------------------------------FOOTNOTES------------------------------------
\medskip
\begin{footnotesize}

\vspace{0.1cm}
\noindent \textbf{Funding Information}: 
%\noindent \textbf{Acknowledgments}: 
We acknowledge support from National Natural Science Foundation of China (62375251) and National Key R\&D Program of China (Grant No. 2020YFB2205800). This work was partially carried out at the USTC Center for Micro and Nanoscale Research and Fabrication.

\vspace{0.1cm}
\noindent \textbf{Author contributions}: 
S.Z. designed, fabricated and characterized the samples. 
S.Z. performed optical measurements. 
S.Z., L.G. and X.L. performed the numerical simulations. 
S.Z., B.S., Z.C., J.Y., J.X., Q.Z., J.L. and A.W. discussed the results. 
S.Z., B.S., J.L. and A.W. analysed the data and finalized the manuscript. 
All authors commented on the manuscript. 
A.W. supervised the project and advised on all efforts.

\vspace{0.1cm}
\noindent \textbf{Conflict of interest}: 
The authors declare no competing interests.

\vspace{0.1cm}
\noindent \textbf{Data Availability Statement}: 
The datasets are available from the corresponding author (A.W., atwang@ustc.edu.cn) upon request.

\vspace{0.1cm}
\noindent \textbf{Supplementary Information}: 
Supplementary Information is available in the journal submission.

\end{footnotesize}

%------------------------------------REFERENCES------------------------------------
\bibliographystyle{apsrev4-1}
\bibliography{references}
\end{document}